\documentclass[a4paper]{jpconf}
\usepackage{graphicx}
\usepackage{wrapfig}

\begin{document}

\title{Annual variations of the $^{214}$Po, $^{213}$Po and $^{212}$Po half-life values}

\author
{
E.N.Alexeev$^1$, A.M.~Gangapshev$^{1,3}$, Yu.M.~Gavrilyuk$^{1}$, A.M.~Gezhaev$^{1}$, V.V.~Kazalov$^{1}$, V.V.~Kuzminov$^{1,3}$, \\ S.I.~Panasenko$^{2}$, O.D.~Petrenko$^{2}$,
S.S.~Ratkevich$^{2}$
}
\address{$^1$ Baksan Neutrino Observatory INR RAS, Russia}
\address{$^2$ V.N.Karazin Kharkiv National University, Ukraine}
\address{$^3$ H.M.Berbekov Kabardino-Balkarian State University, Russia}

\begin{abstract}
Results of a comparative analysis of the $^{214}$Po ($T_{1/2}= 163.47\pm0.03$ $\mu$s), $^{213}$Po ($T_{1/2}=3.705 \pm 0.001$ $\mu$s) and $^{212}$Po ($T_{1/2}=294.09\pm0.07$ ns) half-life annular variation parameters are presented.
It is shown that two independent sequential sets of the $^{214}$Po $\tau$-values $(\tau\equiv T_{1/2})$ obtained in the spaced laboratories can be described by sinusoidal functions.
The sinusoid curve with amplitude $A=(5.0 \pm1.5) \cdot 10^{-4}$, period $\omega=(365\pm 8)$ days, and phase $\phi=(170 \pm 7)$ days approximates the set of $^{214}$Po $\tau$
values obtained at BNO INR RAS during the $\sim$973 days starting on January 4, 2012.
The function approximates a set of $\tau$-values with a time duration of $\sim1460$ days obtained at the KhNU has an amplitude $A=(4.9\pm1.8)\cdot10^{-4}$, a period $\omega= (377\pm13)$ days and a phase $\phi=(77\pm10)$ days. The $^{213}$Po $\tau$-value set with a time duration of $\sim1700$ days can be described by a sinusoidal function with an amplitude $A=(3.9\pm1.2)\cdot10^{-4}$, a period $\omega= (370\pm13)$ days and a phase $\phi=(130\pm9)$ days. The $^{212}$Po $\tau$-value set with a time duration of $\sim670$ days can be described by a sinusoidal function with an amplitude $A=(7.5\pm1.6)\cdot10^{-4}$, a period $\omega= (375\pm13)$ days and a phase $\phi=(40\pm10)$ days.
\end{abstract}

\section{Introduction}

Experimental investigations of a time behavior of the $^{214}$Po-nuclear half-live $(\tau)$ were started at the underground low-background conditions of the Baksan Neutrino Observatory of the INR RAS in April of 2008 year. Results of preliminary measurements carried out during April 2008 - February 2011 showed that amplitude of possible annular variation of the $^{214}$Po half-live does not exceed 0.2\%
of the $\tau$-value averaged at $\sim3$ years \cite{r1}. Measurements with the improved up to $\sim0.02$\%
annular variation sensitivity were continued from the 4 January 2012 year at the modernized set-ups TAU-1 and TAU-2. Annular variation with an amplitude of $A=(8.9\pm2.3)\cdot10^{-4}$,
solar-daily, lunar-daily and sidereal-daily variations with amplitudes $A_S=(7.5\pm1.2)\cdot10^{-4}$, $A_L= (6.9\pm2.0)\cdot10^{-4}$ and $A_{St}=(7.2\pm1.2)\cdot10^{-4}$ \cite{r2,r3} consequently were obtained as a result of a treatment of the $\tau$-values time data sets collected at a period October, 2012 - May, 2015. Half-life of the $^{214}$Po  averaged at 973 days  was found equal to
$T_{1/2}=163.47 \pm 0.03$ $\mu$s.
The work with the $^{214}$Po was relocated from the BNO INR RAS (BNO) to the V.N.Karazin Kharkiv National Unuversity (KhNU).
Measurements were started in December 2015 in the basement of the ground building and are continuing to the present.

Investigations of a time behavior of the $^{213}$Po $\tau$-value using of the TAU-3 installation were started at the BNO in July, 2015. Solar-daily, lunar-daily and sidereal-daily variations with amplitudes $A_S=(5.3\pm1.1)\cdot10^{-4}$, $A_L= (4.8\pm2.1)\cdot10^{-4}$ and $A_{St} =(4.2\pm1.7)\cdot10^{-4}$ \cite{r4} consequently were obtained as a result of a treatment of a data sets collected at a time period (July, 2015 - March, 2017). Half-life of the $^{213}$Po  averaged at 622 days  was found equal to $\tau =3.705\pm0.001$ $\mu$s. The long-time trend embarrasses search for the annular variation was obtained in the data.
The trend shape and annular variation with amplitude $A=(3.2\pm0.4)\cdot10^{-4}$ were found as a result of processing data collected over 1177 days in the period from July 2015 to September 2018 \cite{r5}.

Half-life values ($\tau$) of the examined isotopes are defined as a result of an analysis of decay curves constructed from a set of investigated separate nucleus lifetime values. Delays between the moment of nucleus birth (electron from a decay of $^{214}$Bi($^{213}$Bi)+gamma-quantum ) and the moment of its decay (alpha-particle from a decay of $^{214}$Po($^{214}$Po)) are measured to determine a lifetime value. The useful events identified by a presence of triple pulses one of which is delayed relative to prompt coincides pair pulses. Time sequences of  $\tau$-values with a different time steps are the objects of a subsequent analysis.

Investigations of a time behavior of the $^{212}$Po $\tau$-value were started at the BNO in
July, 2018. A description of the set-up and first result are presented in the Ref.~\cite{r6}. Isotope $^{212}$Po is born
in a $^{232}$Th decay chain after the  $^{212}$Bi beta-decay:
$...\Rightarrow^{212}$Bi[$T_{1/2}$=60.5 min; $\beta$(64\%)
+ $\alpha$(36\%)]
$\Rightarrow$$^{212}$Po($T_{1/2}=2.9\cdot10^{-7}$ s) + $^{208}$Tl($T_{1/2}=3.05$ min) $\Rightarrow$ $^{208}$Pb(stab.) \cite{r7}.
$\beta$-decay of the $^{212}$Bi goes to the ground level in 75.5\%
of occurrences and is not accompanied by the gamma quantum emission.
Emission of a single or cascade gamma-quanta occurs only in 23.8\% of beta decays.
Therefore, the installation efficiency will be 10-15\%
for triple [$(\gamma \otimes \beta) \otimes \alpha $]-coincidences when the efficiency of gamma-registration is taken into account.
This essentially decreases the total efficiency of the installation.
It was decided to use double delayed coincidences [$\beta\otimes\alpha$] to improve the efficiency of the selection of useful events.
It is possible due to very short $^{212}$Po half-life and low influence of a detector background on an accuracy of a $\tau$ definition due to this.
Massive NaI(Tl) detectors used usually for the gamma-registration was not included in the new set-ups. A mass of a low background shield surrounded the detector was essentially reduced because of it. Two short testing
measurement series were carried out in the ground level laboratory (680 hours) and in the underground one (564 hours).

Average over the entire measurement time value of the $^{212}$Po half-life measured in the ground level laboratory was found to be $\tau=294.09\pm0.07$ ns, and the one measured underground had almost the same value $\tau=294.07\pm0.08$ ns \cite{r8}. Solar-daily variations with amplitudes $A_S=(11.7\pm5.2)\cdot10^{-4}$ and
$A_S=(7.5\pm4.1)\cdot10^{-4}$ were obtained in the ground and underground time series of the $\tau$-values consequently.  The measurements with this isotope are continuing at the present time. The data analysis results collected to February 2020 for these three isotopes have presented at this work.

\section{Measured results}

\hspace{1pc} {\bf $^{214}$Po}

The entire data array accumulated at the BNO during measurements with $^{214}$Po at the TAU-2 facility was re-processed without using the ``internal moving average'' method described in the Ref. \cite {r3}.
This method was used earlier for a search for long-time variations with a required annular frequency and did not allow observing variations with different frequencies. A decay curve at a specified time interval in a new approach was constructed with the birth-decay $^{214}$Po delays contained in a one or several serial computer files completed in the ``on line'' mode.
The duration of a separate file was not set strictly but depended on the periodicity of stops for calibrations and equipment repairs. The file accumulation time was $\sim$14-30 days.
The file with the 14 duration contains initially $\sim1.4\cdot10^7$ events if a count rate was equal to $\sim12$ s$^{-1}$.
\begin{wrapfigure}[30]{0.5}{0.5\textwidth} \vspace{-1pc}
\includegraphics[width=19.0pc,angle=0]{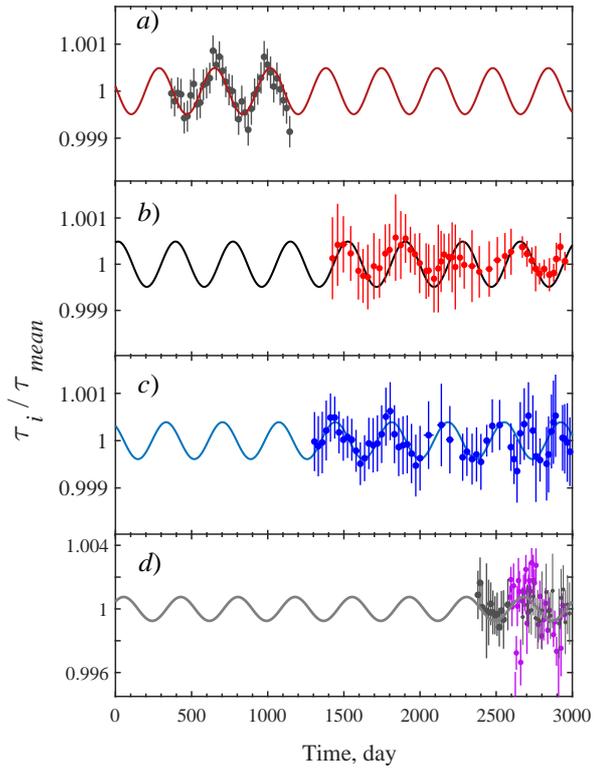}
\vspace{-2.0pc}
\caption{\label{fig1}
{\small
Time variations of the half-life values of the $^{214}$Po, $^{213}$Po and $^{212}$Po isotopes. $a$) $^{214}$Po TAU-2 set-up (BNO); $b$) $^{214}$Po TAU-KHR set-up (KhNU); $c$) $^{213}$Po TAU-3 set-up; $d$) $^{212}$Po  TAU-4 and TAU4d set-ups.
}}
\end{wrapfigure}
The events with amplitudes correspond to ranges of the preset thresholds in amplitude $(\alpha,\beta,\gamma)$-spectra were used to construct the decay curves. $\gamma$-quanta registered by two massive scintillation NaI(Tl)-detectors.
$\beta$- and $\alpha$-particles registered by two plastic scintillator disks.
The mother $^{226}$Ra-source placed between discs inside a thin bag made of 2.5 $\mu$m polyethylene terephthalate (PET) film. The last detector seats between NaI(Tl) crystals. A decay curve approximated by a function $y=A\cdot exp(-t/\tau)+B$, where $B$ is a random coincidence background and $t$ is a time between a birth and decay of $^{214}$Po nucleus. All mathematical operations with data and approximations are performed in the MATLAB software package.

When examining the results, it was found that the error in determining the $\tau$-value was $\pm0.2$\%,
and the spread of values did not allow to extract any variation without additional averaging and smoothing.
Preference was given to smoothing the array of birth-decay time delays in a given measurement time interval before an exponential fit to a decay curve.
We have used the Hamming filter, which allows us to smooth out high-frequency outliers in the data series \cite{r9}.
As a consequence, we can obtain the parameters of the decay curve with high accuracy.
The Hamming window width was chosen close to $0.5 \times T_{1/2}$ for a digital time delays series.
The start and end of the numerical delay data passed through the filter were discarded to remove edge effects and eliminate distortion during the exponential fit to the decay curve.
Preliminary testing of this method showed that the result of fitting the initial data and the data passed through the Hamming filter gives the exponent parameters an order of magnitude more accurate.
So for the decay curve collected during the exposure of 28 days, the half-life was $\tau = 163.511 \pm 0.415$ and $\tau = 163.518 \pm 0.048$ $\mu$c for the original and smoothed data, respectively.
The error has decreased nine times, and the value of $\tau$ has changed by 0.005\%.
The window width kept the same during the processing of the entire data set.

A relative changes of the $\tau$-value is of interest at the presented investigation than to make possible a comparison of the parameter obtained at different set-ups or with different isotopes a $\tau$-value in a particular point normalized on the unity by a division on the  $\tau$-value averaged through all data set.
A time dependence of $^{214}$Po half-life at the TAU-2 set-up has shown on the graph ({\it a}) of Fig.\ref{fig1}.
The days since 1 Jannuary, 2012 apply on the
x-axis. Long-time variation is clearly seen on the picture.
This cycling were approximated by sine function $y=A\cdot sin(2\pi/\omega(t+\phi))$, where $A=(5.0\pm1.5)\cdot10^{-4}$  is an amplitude, $\omega=(365\pm8)$ days is a period and $\phi=(170\pm7)$ days is a phase. Here and after phase is a time moment preceding to the 1 January, 2012  were the function equal to zero and starts to rise with the argument growth.

As already mentioned, measurements with the isotope $^{214}$Po continued at the KhNU (since December 2015).
Registration of $\alpha$-particles from the $^{214}$Po decays in the TAU-KHR set-up is carried out by the two silicon detectors with a source located between them.
The source is a radium salt sealed between two layers of 2.5 $\mu$m thick PET film, glued along the edge with epoxy.
The semiconductor detectors assembly with the source is placed in a container $d = 40$ mm, $h=20$ mm. The container is located on the surface of the NaI(Tl) detector $d=160$ mm, $h=100$ mm, designed to register $\gamma$-quanta from the decay of $^{214}$Bi. The installation is housed in a shield made of 20 cm Fe. Since the thickness of the active silicon surface barrier detectors layer is small, the energy released from the $^{214}$Bi decay electron crossing this layer is also small. Much of these events are lost in preamplifier noise.
A count rate of useful events with the triple [($\gamma \otimes \beta) \otimes \alpha$] coincidence considerably decreased in comparison with the TAU-3 set-up and is equal to $\sim3$ s$^{-1}$. A treatment of a data was the same as described above. The result is shown on the graph ($b$) of  Fig.~\ref{fig1}. As it seen, a long-time variation of the $^{214}$Po half-life takes place in the new conditions too. Sine function approximates the TAU-KHR data has parameters $A=(4.9\pm1.8)\cdot10^{-4}$, $\omega=(377\pm13)$ days and $\phi=(77\pm10)$ days.
\\

{\bf $^{213}$Po}

\begin{wrapfigure}[14]{0.5}{0.5\textwidth} \vspace{-2pc}
\includegraphics[width=17.0pc,angle=0]{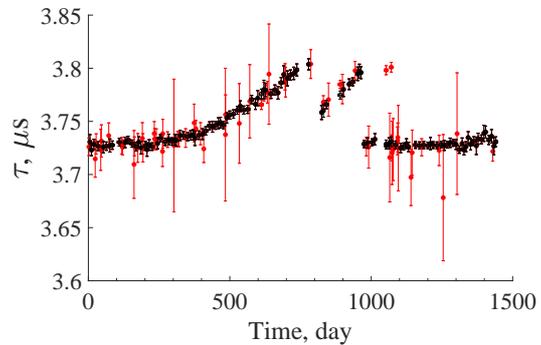} \vspace{-1pc}
\caption{\label{fig2} {\small Time dependence of the $^{213}$Po half-life values averaged on integer number of files (July 2015 -- October 2019; 1555 days).
}}
\end{wrapfigure}
Measurements with the $^{213}$Po carry out on the same installation which was used for the $^{214}$Po investigation. A bag with the $^{226}$Ra source was changed by a bag with the $^{229}$Th mother source. The new version of a set-up was named TAU-3. A recording rate of useful events equal to
$\sim$18 s$^{-1}$. As it was mentioned, a large-scale noncyclic trend was found in a process of the analysis of the averaged at one week $^{213}$Po $\tau$-values computed from the experimental data.  A $\tau$-value starts to grow after $\sim$42 weeks measurements and increased at $\sim$0.6\%
to the end of $\sim$742 days interval in a comparison with a value averaged at the first 320 days. A $\tau$-value returned by a jump to the initial level at the 182 weeks (9 January, 2019) and keeps the value to the end of a processed interval. The reasons of such behavior are not determined explicitly at a present time. It was no any interference into the set-up work except routine testing procedures. The equipment is not changed.
The data from the TAU-3 set-up was processed as mentioned above to take a notion about possible long-time variations at a large-scale noncyclic trend background. A time dependency of the $^{213}$Po half-life values averaged at $\sim$14-30 day time intervals is shown on Fig.~\ref{fig2}.
It is seen that a $\tau$-value have noncyclic one-sided deviations.

Similar trend is absent in the data of TAU-KHR set-up which works simultaneously with TAU-3.
It allows considering that the trend does not connected with a source of periodic variations and can be eliminated from a data set.
Half-life value averaged over the 0-320-day interval was subtracted from the entire data series. A one-side deviation on the interval 300-800 days was approximated by a second order polynomial and by a straight line on the interval 801-1000 days. Approximations were subtracted from the data set. The result normalized to unity shown of graph. $c$) of Fig.\ref{fig1}. Long-time cyclic variations are visible on the graph. Sine function approximates the TAU-3 data has parameters $A=(3.9\pm1.2)\cdot10^{-4}$, $\omega=(370\pm13)$ days and $\phi=(130\pm9)$ days.
\\

{\bf $^{212}$Po}

Assumptions that viewed variations of the tested isotopes half-lives could be connected with changing of the installation working characteristics caused by environmental factors variations
were tested during the whole period of the measurements. None of the checks have confirmed such influence so far. Though, such probability could not be excluded totally because of existence of a large quantity of active environmental factors (pressure, air humidity and temperature, layer of air ionization, earth magnetic field and others) and difficulties of its detection with the required level of  sensitivity at a presence level of useful event count rates. It was found with a start of the $^{212}$Po half-life measurement at the TAU-4 set-up and using of the Hamming filtering for a decay curve smoothing \cite{r6} that deviation of $\tau$-values averaged at 2-4 hours from the one averaged at 1 week essentially exceeds  the statistical errors. Such behavior of the investigated value at the 1 day scale showed a possibility to test a belonging of different installation components to an appearing of the variations. The TAU-4d (double) set-up was created for this investigation. A light from the one active element (scintillator + source) registered by two independent circuits consist of (PMT ET9203B + high voltage source + driver circuit + digital oscilloscope).

\begin{figure}[h]
\includegraphics[width=18pc]{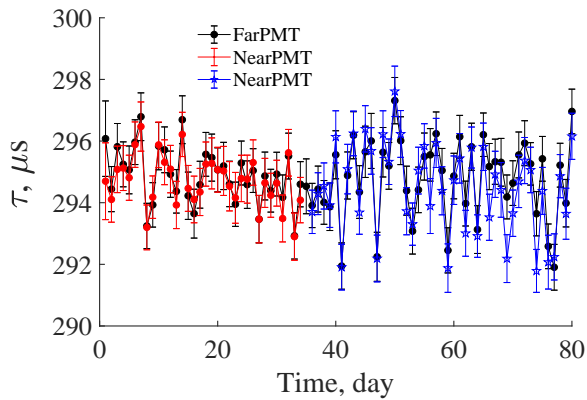} \hspace{1pc}
\begin{minipage}[b]{18pc}
\caption{\label{fig3}
{\small
Time dependences of the $^{212}$Po half-life values averaged at one day for the two TAU-4d channels. A start of the measurement is 31 January, 2019. 1) 0-34 days, DO LA-1nUSB with the 500 MHz sampling frequency installed in the both channels. DO recording starts by a single DC connected to the left PMT; 2) 35-55 days, DO LA-1nUSB (500 MHz) installed in one channel and DO NI-5124 (200 MHz) installed in the other one. DO recording starts by a common DC; 3) 56-80 days, one is LA-1nUSB and second is NI-5124. Two independent DC.
}}
\end{minipage}
\end{figure}

It was assumed on a sample of TAU-4 set-up with one circuit (PMT Hamamatsu R12669 SEL + driver circuit (DC) + digital oscilloscope (DO)) that variations of the half-life value will be independent in the TAU-4d different channels in a case of one of the listed block instability. The variations will correlate between themselves and with the half-life variation obtained at the independent TAU-4 set-up if they are caused by the influence an unknown factor on the one of the set-ups block. The results of the approximated $\tau$-values for the left (l) and right (r) channels of the TAU-4d setup are shown in Fig.~\ref{fig3} by black circles and (red+blue) dots, respectively.
Different versions of the channel configuration and ways of start DO
recording were tested in a process of work with the TAU-4d. There were tested versions: 1) 0-34 days, DO LA-1nUSB with the 500 MHz sampling frequency installed in the both channels. DO recording starts by a single DC connected to the left PMT; 2) 35-55 days, DO LA-1nUSB (500 MHz) installed in one channel and DO NI-5124 (200 MHz) installed in the other one. DO recording starts by a common DC; 3) 56-80 days, one is LA-1nUSB and second is NI-5124. Two independent DC. It is seen from a comparison of both channel graphs that their forms practically coincide independently on the measurement conditions.

\begin{figure}[ht]
\begin{minipage}{17pc}
\includegraphics[width=17pc]{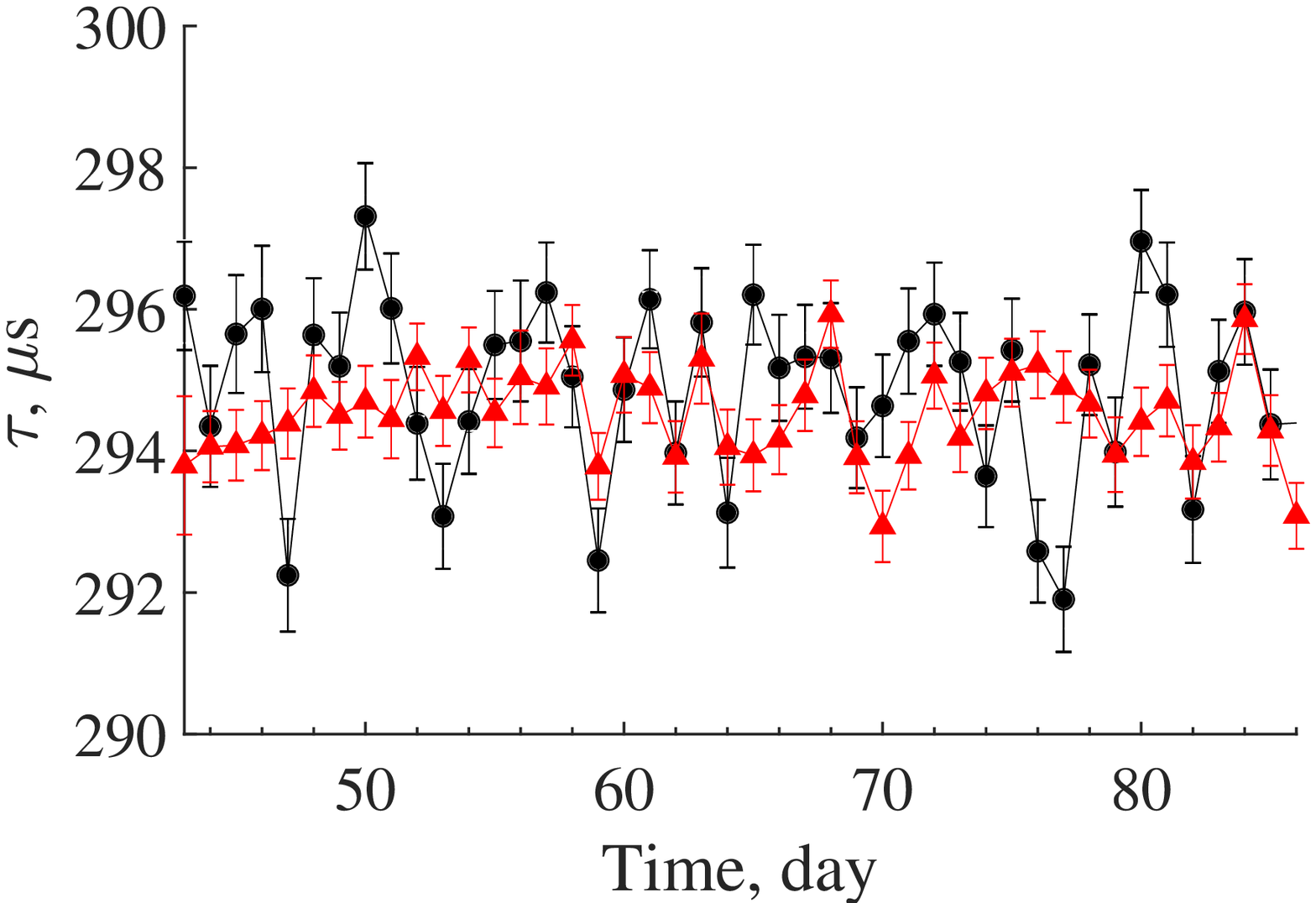}\vspace{0pc}%
\caption{\label{fig4} {\small (Time dependencies of the $^{212}$Po half-life values averaged at one day for the right channel of the TAU-4d (PMT-r) (black) and for the TAU-4 (red). A start of the measurement is 43 day after 31 January, 2019.
}}
\end{minipage}%
\hspace{2pc}%
\begin{minipage}{17pc}
\includegraphics[width=18pc]{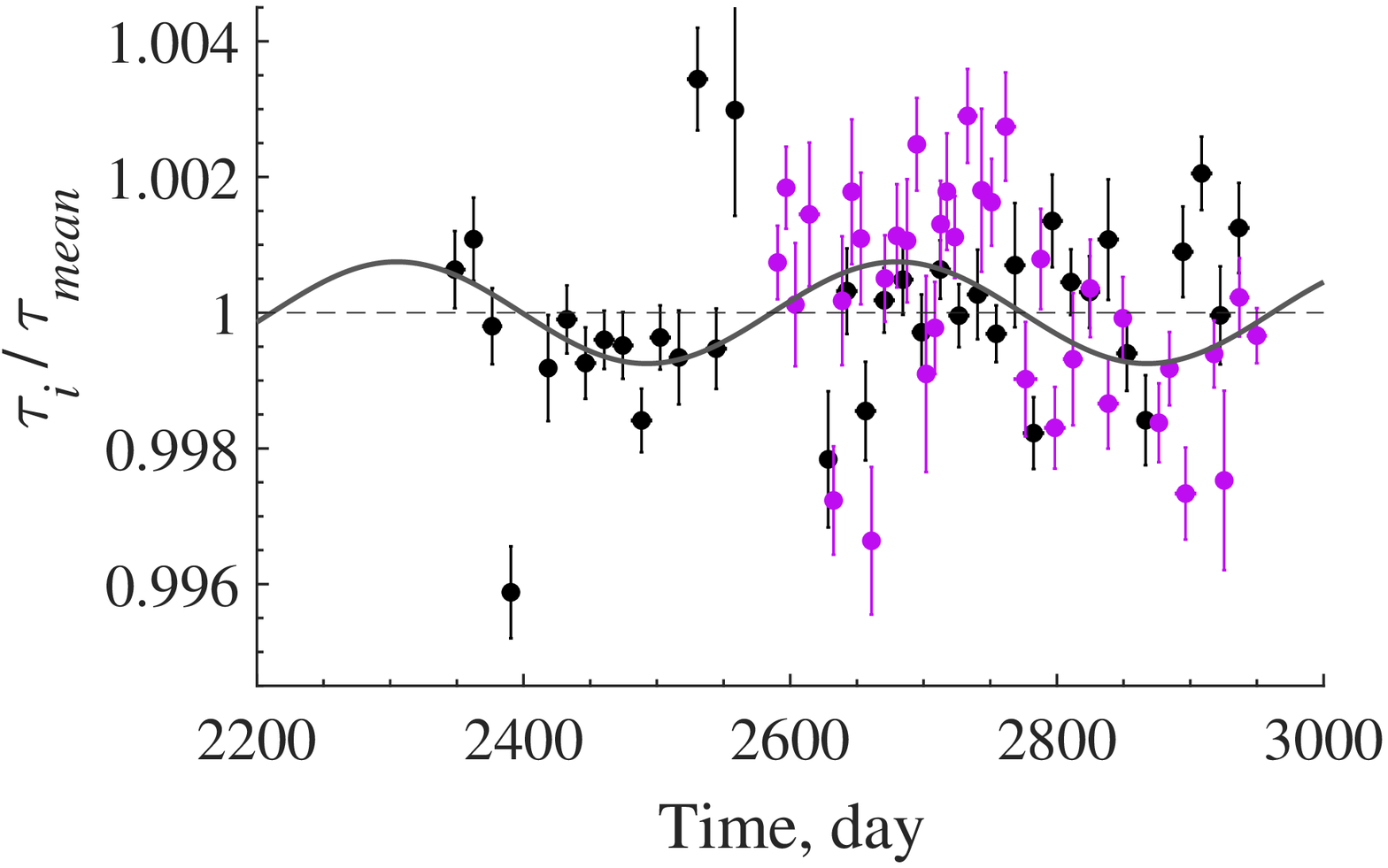}\vspace{0pc}%
\caption{\label{fig5}Time dependencies of the $^{212}$Po half-life values at the TAU-4 set-up (black) and at the TAU-4d, right channel (pink).  Approximation functions is $y=7.5\cdot10^{-4}\cdot sin[2\pi/375\cdot(t+40)]$.}
\end{minipage}
\end{figure}

Analogous results for the right cannel of the TAU-4d (black) and for the TAU-4 are shown on Fig.~\ref{fig4}.
The channels are fully identical except the PMTs. It is seen that dependencies strongly different.
The fact that count rate on the TAU-4d set-up is 2.5 times less the one of the TAU-4 should be taken into account. A situation does not changed at changings of distances between set-ups and their
orientations. The conclusion should be made that short period noise-type deviations of the $^{212}$Po half-life values in the TAU-4d and TAU-4 set-ups are connected with the active element (plastic scintillator + radioactive source).

Schedule of TAU-4d and TAU-4 set-ups operations noticeably differs in the start time moments and in the stop numbers and its durations. They completed one for another in the $\sim$1.7 years (30-May-2018 -- 20-Feb-2020) interval of the measurement time. The results of a data processing according to the method described above are normalized to the unity and presented on the Fig.~\ref{fig5} (TAU-4 - black, TAU-4d (PMTr) - pink) and Fig.\ref{fig1}, graph $d$). Sine function approximates the data (black line) has parameters $A=(7.5\pm1.6)\cdot10^{-4}$, $\omega=(375\pm13)$ days and $\phi=(40\pm10)$ days.

\section{Result discussion }

It is follows from the reasons listed above that active element composed of the plastic scintillator and the radioactive source is the most probable source of the variations if a PMT does not enter own deposit to this process. An absence of a such influence could be prove on a sample of
measurements with the $^{214}$Po and $^{213}$Po which were performed on the same set-up. It is required to change average delays between responses from $\beta$- and $\alpha$-particles to appear a variation of the decay curve parameter $\tau$.
In other words, an unknown assumed time characteristic of a PMT should change. A possible influence of outer factors should cause proportional changes of this hypothetic characteristic. The absolute values of possible variation of a PMT parameter are identical in the both isotope measurements. Amplitude of a relative changing of the $^{213}$Po $\tau$-values will increase at $\sim40$ times in accordance with a $\tau$-value decreasing because of it is proportional to the absolute amplitude of an average delay deviation at the first approximation. It does not happen as it see from the Fig.~\ref{fig1}. The conclusion that PMTs just as the other blocks don't influence on a $\tau$-value variation could be done from this consideration.

A conclusion that half-life of the $^{214}$Po feels an annual sine variation independently on the set-up location, environmental conditions, the detector type and upper level of radioactive background rejection could be done from a comparison of graphs $a$) and $b$) on the Fig.~\ref{fig1}.

Half-life of the $^{213}$Po feels an annual variation also as it see from the graph $c$) of Fig.~\ref{fig1}. A sine amplitude defined for the total data set somewhat less the one defined for the first half of the data set. It could be connected with insufficiently accurate choice of curves approximated the trend.
Amplitude of the $^{213}$Po at the first part of a data set is close to the $^{214}$Po one.

Duration of the $^{212}$Po half-life data set is not so long but the presented data could be approximated by a sine function also as it seen from the graph $d$), Fig.~\ref{fig1} and Fig.~\ref{fig5}.
Sine wave amplitude is slightly higher than a  $^{214}$Po amplitude.

\section{Summary and conclusion}

Conclusion that half-life values of the $^{214}$Po, $^{213}$Po and $^{212}$Po isotopes are feel annular variations with similar amplitudes could be done on a base of a comparison of graphs on the Fig.~\ref{fig1}.
It is shown that two independent sequential sets of the $^{214}$Po $\tau$-values ($\tau \equiv T_{1/2}$) obtained in the spaced laboratories can be described by sinusoidal functions. A sine function approximates a set of the $^{214}$Po $\tau$-values with a time duration of $\sim973$ days obtained at the BNO has an amplitude $A=(5.0\pm 1.5)\cdot 10^{-4}$, a period $\omega=(365 \pm 8)$ days and a phase $\varphi=(170 \pm 7)$ days relative to the 1st January, 2012 year. The function approximates a set of $\tau$-values with a time duration of ~1460 days obtained at the KhNU has an amplitude $A=(4.9 \pm 1.8)\cdot 10^{-4}$, a period $\omega = (377 \pm 13)$ days and a phase $\phi=(77 \pm 10)$ days. The $^{213}$Po $\tau$-value set with a time duration of $\sim1700$ days can be described by a sinusoidal function with an amplitude $A=(3.9 \pm 1.2) \cdot 10^{-4}$, a period $\omega = (370 \pm 13)$ days and a phase $\phi=(130 \pm 9)$ days. The $^{212}$Po $\tau$-value set with a time duration of $\sim670$ days can be described by a sinusoidal function with an amplitude $A= (7.5 \pm 1.6) \cdot 10^{-4}$, a period $\omega = (375 \pm 13)$ days and a phase $\phi=(40 \pm 10)$ days.

Observed differences of the phases of the approximated sine functions have not any unambiguous explanations and define necessity of further investigations. These differences are complicate a search of the common factors caused the viewed variations. None of environmental factors examined up to now (pressure, air humidity and temperature, layer of air ionization, earth magnetic field variations, set-up instability) do not show correlations with the observed $^{214}$Po, $^{213}$Po and $^{212}$Po half-life values variations. Investigations are continuing.

The authors are expressing their gratitude to V.I.Volchenko and A.F.Yanin for a development and fabrication of the Driver Circuits.

The work was made under the financial support of the ``Physics of hadrons, leptons, Higgs bosons and dark matter particles'' Program of the RAS.


\end{document}